\documentclass[11pt,a4paper]{article}
\usepackage[utf8]{inputenc}
\usepackage{anyfontsize}
\usepackage{tikz}
\usepackage{mathtools, nccmath}
\usepackage[edges]{forest}
\usepackage{bm}
\usetikzlibrary{arrows.meta}
\fontfamily{cmr}
\usetikzlibrary{matrix,shapes,arrows,positioning,chains,shapes.geometric}
\usepackage[margin=1in]{geometry}
\usepackage{amsmath, amsthm, amssymb,amsfonts}
\usepackage{graphicx}
\usepackage{tcolorbox}
\usepackage{enumitem}
\usepackage{cite}
\usepackage{hyperref}
\begin{document}
\title{\Large\bf{On the Method of Brackets}
}

\author{\bf B. Ananthanarayan$^a$, Sumit Banik$^a$, Samuel Friot$^{b,c}$ and Tanay Pathak$^a$}
\date{}
\maketitle
\begin{center}
{$^a$ Centre for High Energy Physics \\Indian Institute of Science, Bangalore-560012, Karnataka, India}\\[0.5cm]
{$^b$ Institut de Physique Nucl\'eaire d’Orsay \\
Universit\'e Paris-Saclay, CNRS/IN2P3, IJCLab, 91405 Orsay, France
}\\[0.5cm]
{$^c$ Institut de Physique Nucléaire de Lyon \\ Univ Lyon, Universit\'e Claude Bernard Lyon 1, CNRS/IN2P3, IP2I Lyon, France}
\end{center}

\begin{abstract}

The Method of Brackets (MoB) is a technique used to compute definite integrals, that {has its origin} in the {negative dimensional integration method}. It was originally proposed for the evaluation of Feynman integrals for which, when {applicable, it gives} the results in terms of combinations of (multiple) series. We focus here on some of the {limitations} of MoB and {address them by studying the} Mellin-Barnes (MB) representation technique. {There has been significant process recently in the study of the latter} due to the development of a new computational approach based on conic hulls (see Phys. Rev. Lett. 127, 151601 (2021)). 
The comparison between the two methods helps to {understand} the limitations of the MoB, in particular when termwise divergent series appear. {As a consequence, the MB technique is found to be} superior over MoB for two major reasons: 1. the selection of the sets of series that form a series representation for a given integral follows, in the MB approach, from specific intersections of conic hulls, which, in contrast to MoB, does not need any convergence analysis of the involved series, and 2. {MB can be used to evaluate resonant (i.e. logarithmic) cases where MoB fails due to the appearance of termwise divergent series.}
Furthermore, we show that the recently added Rule 5 of MoB naturally emerges as a consequence of the residue theorem in the context of MB.

\end{abstract}

\section{Introduction}
The evaluation of Feynman integrals, which  
arise in the perturbative framework of quantum field theory, is of crucial importance for precision studies of the observables in the Standard Model. In this respect, numerical and analytic methods continue to be developed by the particle physics community. Well-known methods like integration-by-parts, differential equations, Mellin-Barnes representation, sector decomposition, the method of expansion by regions or subgraphs, etc. (see \cite{Smirnov} for a review of such techniques) have proven their usefulness during the last decades.

 We focus here on {yet another method} which is the Method of Brackets (MoB) \cite{Gonzalez:2010uz,gonzalez2010definite1,gonzalez2010method2,Gonzalez:2007ry,GONZALEZ2015214,gonzalez2016pochhammer,gonzalez2017extension,GonzalezJiuMoll}. It is built on a set of heuristic rules, originally formulated to evaluate Feynman integrals\cite{Gonzalez:2010uz,gonzalez2017moments,gonzalez2019analytic}, but which is also useful for the computation of certain definite integrals \cite{gonzalez2010definite1,gonzalez2010method2,gonzalez2014evaluation,Ananthanarayan:2019hia}. The MoB, which is based on a generalization of Ramanujan's Master Theorem \cite{Amdeberhan2012, Hardy}, was proposed as an optimized version of the Negative Dimensional Integration Method (NDIM), the latter having been first introduced in  \cite{Halliday:1987an,dunne1987negative} and applied to Feynman integrals in \cite{suzuki2004evaluating,suzuki2002massless,suzuki2003general,suzuki1999feynman,suzuki2000negative,PhysRevD.58.047701,anastasiou2000application,anastasiou2000scalar,DelDuca:2009ac}. The MoB simplifies the problem of solving an  integral into finding the solutions of a system of linear equations and often provides the answer as a set of multi-fold basis series (which, when convergent, are in most cases multivariable hypergeometric series), some appropriate linear combinations of which give the series representations of the integral. Although this method is fairly easy to apply, it is not very useful in many cases, as finding the correct linear combinations is usually the hardest step. Indeed, this step requires a complete convergence analysis of each series (obtained from the MoB) which is poorly understood for higher-fold hypergeometric series. Another important drawback of the MoB is the frequent appearance of termwise divergent series which, when discarded (as dictated by one of the heuristic rules of MoB), leads to incomplete results. This latter point is linked to the important limitation of MoB which cannot deal with resonant (i.e. logarithmic) cases. 

It is one of the objectives of this work to explicitly point out and discuss these limitations. We propose to perform this exercise in the context of a comparison of MoB against the Mellin-Barnes (MB) approach. It has been suggested several times in the past to study the links between these two methods and even to mix them \cite{gonzalez2017moments,Suzuki:2003jn,Prausa:2017frh} (see, by the way, the recent work \cite{Gonzalez:2021vqh} where MoB is used to compute some one-fold MB integrals). However, and this is the second aim of the present work, we will show that the significant progress that has been made in MB theory \cite{ananthanarayan2020multiple}, by solving the important issue of deriving series representations of MB integrals of a very general form, shows the superiority of MB over MoB, since the former does not have any of the drawbacks, mentioned above, of the latter. Moreover, this new MB computational method has been fully automatized in the \texttt{MBConicHulls.wl} \textit{Mathematica} package \cite{ananthanarayan2020multiple}, making it straightforward to apply, which is not the case of MoB.
This MB technique, based on conic hulls, was developed by three of the authors to express $N$-fold MB integrals in terms of multivariable series representations. It can be of great interest in the many fields where multiple MB integrals appear: obviously in physics, but also in mathematics as for instance in the theory of hypergeometric functions of several variables \cite{srivastava1985multiple}, or in asymptotic theory \cite{Paris&Kaminsky}. In quantum field theory, this conic hulls MB method (CHMB) is expected to impact Feynman integrals computations as well,
since it is well-known that they can be expressed in terms of MB integrals \cite{Smirnov}. 

One of the important advantages of the CHMB, in comparison to some earlier attempts to solve multiple MB integrals, is that one is now able to derive the various series solutions without any convergence analysis, and also without performing the MB integrations iteratively. These series representations can be made of combinations of convergent multiple series and/or divergent asymptotic ones, depending on the form of the MB integrand.
An important result, in the so-called degenerate case \cite{ananthanarayan2020multiple} where only convergent series appear in the series representations, is that one can in general find a master series for each series representations, which facilitates the numerical checks and convergence analysis when the latter are needed. At last, the CHMB method can also handle resonant (i.e. logarithmic) cases, where poles of higher-order in the MB integrand are involved. Although this requires to apply multivariate residues theory in a less simple way than for the nonresonant case, the corresponding calculational procedure has also been automatized in the \texttt{MBConicHulls.wl} package.

In this paper, we establish a connection between the MoB and CHMB in the nonresonant case (see Section \ref{Equivalence_Section}). Based on 
this connection, we qualitatively explain the origin of termwise divergent series in the MoB,
which in the CHMB case manifests itself
as the presence of higher-order poles in the MB representation's integrand. Let us recall that one of the heuristic rules of MoB (Rule 4) asks that these divergent series are simply discarded. This, as we will show, leads to incomplete results. Therefore, we use this connection to conclude that not only
(a) the CHMB can solve all integrals which are solvable by the MoB, but also that (b) resonant cases, which can be dealt with CHMB,
are not computable with MOB which gives incomplete results due to the presence of termwise divergent series. To make our discussion explicit, 
a suitable Feynman integral is treated and solved. This example exhibits all the essential points that we wish to discuss.  

 The outline of this paper is as follows: In Section \ref{Method_Intro_Section}, we briefly describe the formalism of the CHMB and the MoB. In Section \ref{Equivalence_Section}, a detailed discussion is made to establish a connection between the methods.
 This connection is then used in Section \ref{Limitation_Section} to explain the origin of divergent series in the MoB. Section \ref{Box_Diagonal_Solve} is devoted to discussing a Feynman integral in order to illustrate our discussion of the previous section. The details of the corresponding calculations are shown in a \textit{Mathematica} notebook given as an ancillary file to this paper. Finally, in Section \ref{Conclusion_Section}, we present our concluding remarks.  In Appendix A, we show that Rule 5 of the MoB naturally comes from the residue theorem as applied in CHMB while Appendix B gives the proofs of several identities paving the way of our discussion in Section \ref{Equivalence_Section}.

\section{A brief introduction to the methods}\label{Method_Intro_Section}

We begin with outlining the basic formalism, ideas and steps involved in each of the MoB and the CHMB. In practice, both methods are straightforward to use.
Several worked out examples on the CHMB have been provided by some of us in \cite{Ananthanarayan:2020ncn,Ananthanarayan:2020xpd,ananthanarayan2020multiple},
those arising from the application of MoB can be found in \cite{Gonzalez:2010uz,gonzalez2010definite1,gonzalez2010method2,Gonzalez:2007ry,GONZALEZ2015214,gonzalez2016pochhammer,gonzalez2017extension,GonzalezJiuMoll}. 
As mentioned in the introduction, a fully automatized \textit{Mathematica} package of the CHMB also exists \cite{ananthanarayan2020multiple}, making this method even more accessible to the reader.

Let us begin with the MoB in Section \ref{MoB_Rules} and proceed to CHMB in Section \ref{CHMss}.  

\subsection{Method of Brackets}\label{MoB_Rules}
The MoB is based on a multi-variable generalization of Ramanujan's Master Theorem \cite{Amdeberhan2012,Hardy}. It is comprised of a set of rules that, when applied to the Schwinger parametric representation of a Feynman integral or to certain type of definite integrals, yields a single term solution or a set of basis series. We list below the rules of this method. 
\begin{itemize}
\item \textbf{Rule 1 :} A function $h(x)$ in the integrand should be replaced by its power series expansion of the form 
\begin{equation}
    {
    h(x)=\sum_{n=0}^\infty a_n x^{\alpha n + \beta }
    }
\end{equation}
\item \textbf{Rule 2 :} A multinomial in the integrand should be replaced by its bracket series of the form
\begin{equation}
{
    (a_1 + a_2 +\dotsm+ a_r)^{\alpha}=    \frac{1}{\Gamma(-\alpha)}\, \, \, \, \,\sum_{\mathclap{m_1,\dotsc, m_r}}\phi_{m_1,\dotsc,m_r}a_{1}^{m_1}\dotsm a_{r}^{m_r}\langle -\alpha+m_1+\dotsm+m_r\rangle
    }
\end{equation}
where, the bracket is defined as $\int_0^\infty dx\;x^{\alpha-1}=\left\langle\alpha\right\rangle$.
It is obvious from this definition that the bracket is a divergent integral for all values of ${\alpha}$ but it plays its role when inserted inside an infinite series as
\begin{equation}
       \sum\limits_{m=0}^{\infty} \phi_m
        \left\langle m+\alpha \right\rangle
        \,F(m)
        =
        \int\limits_0^\infty dx\;  x^{\alpha - 1}
        \sum\limits_{m=0}^{\infty} \phi_m \;
        x^{m}\,
        F(m)
        =
        F(-\alpha)\Gamma(\alpha)\label{EqBracket}
      \end{equation}
      where $\phi_m=\frac{(-1)^m}{\Gamma(1+m)}$, $F(m)$ is a formal function of $m$ and, in the last step to derive Eq. (\ref{EqBracket}), we used Ramanujan's Master Theorem.
\item \textbf{Rule 3 :} Once Rules 1 and 2 are applied to each factors in the integrand, one can combine the powers of the integration variable (say $x$) and introduce brackets using the definition of the bracket
\begin{equation}\label{Rule_3}
{
 \int_0^\infty dx\;x^{\textbf{L}-1}=\left\langle\textbf{L}\right\rangle}
\end{equation} 
 where \textbf{L} is a function of the summation indices and other parameters of the integral.
\item \textbf{Rule 4 :} The brackets replaces the integration symbols and converts the integral into a series with brackets in the summand 
\begin{equation}
 I=\sum_{\mathclap{n_1,\dotsc,n_r}} \phi_{n_1,\dotsc,n_r}\,\,f(n_1,\dotsc.,n_r)
                            \langle b_{11}n_{1}+\dotsm+b_{1r}n_{r}+c_{1}\rangle
                               \cdots \langle b_{s 1}n_{1}+\dotsm+b_{s r}n_{r}+c_{s}\rangle
\end{equation}                        
termed as the bracket series, where $f$ is a formal function of the summation indices variables, ${r\geq s}$ and $\phi_{m_1,\dotsc,m_r}\doteq\phi_{m_1}\phi_{m_2}\cdots\phi_{m_r}$. 

The bracket series can be rewritten as
\begin{equation}\label{Bracket_Def}
 I=\sum_{\eta} \phi_\eta\,\,f(\eta)
                            \langle B \eta +C \rangle
                        \end{equation}        
 where the matrices $B$, $\eta$ and $C$ are
\begin{equation}\label{B_Matrix}
B=\left(
\begin{array}{ccccc}
 b_{11} & b_{12} & \cdots & b_{1r} \\
 b_{21} & b_{22} & \cdots & b_{2r} \\
  & & \cdots & \\
 b_{s1} & b_{s2} & \cdots & b_{sr} \\
\end{array}
\right), \hspace{1cm} 
\eta=\left(
\begin{array}{ccccc}
n_1 \\
n_2 \\
. \\
 n_r \\
\end{array}
\right), \hspace{1cm} 
C=\left(
\begin{array}{ccccc}
c_1 \\
c_2 \\
. \\
 c_r \\
\end{array}
\right)
\end{equation} 
and where we used the condensed notations  $\phi_\eta\doteq\phi_{n_1} \, \cdots \, \phi_{n_r}$, $f(\eta)\doteq f(n_1, \cdots\, , n_r)$ and $ \langle B \eta +C \rangle\doteq \langle b_{11}n_{1}+\dotsm+b_{1r}n_{r}+c_{1}\rangle
                               \cdots \langle b_{s 1}n_{1}+\dotsm+b_{s r}n_{r}+c_{s}\rangle$. As in Section \ref{CHMss}, we shall use the notation $B^{\sigma}$ (respectively $B_{\sigma'}$) to denote the matrix constructed from the rows labeled by the set of indices $\sigma$ (respectively the columns labeled by the set of indices $\sigma'$) of $B$.\\\\
Let us now consider the two possible cases of Eq.\eqref{Bracket_Def}.
       \begin{enumerate}[label=(\roman*)]
\item For $\bm{r=s}$, the solution of the integral is a single term
\begin{equation}\label{MoB_Single_term_1}
{
I=\frac{1}{|\mathrm{det} \, B|}f(n^{*}_{1},\dotsc.,n^{*}_r)\Gamma(-n^{*}_{r})\cdots\Gamma(-n^{*}_{1})}
                        \end{equation}
where ${n^{*}_i}$ are the solutions of ${n_i}$ by solving the system of linear equations obtained by setting the argument of each bracket in Eq.\eqref{Bracket_Def} to zero. Thus, in the matrix notation we can write $\eta^*=-B^{-1}\,C$ and rewrite Eq.\eqref{MoB_Single_term_1} as,
\begin{equation}\label{MoB_Single_term}
{
I=\frac{1}{|\mathrm{det} \, B|}f(-B^{-1}\,C)\Gamma(-B^{-1}\,C)}
                        \end{equation}

where, for a given column matrix $X$ whose $i^{th}$ row is $X^i$, we introduce the notation $\Gamma(X)=\prod_{i}\Gamma(X^{i})$. This will be followed for the rest of the paper.
 \item For $\bm{r>s}$, one obtains a set of $r-s$ fold series, which we term as basis series. To get them we need to consider all possible $r-s$ combinations of free summation variables among $r$ summation variables in matrix $\eta$ given in Eq.\eqref{B_Matrix} and associate each choice with a basis series. 
 
Let us take $\eta^{{\sigma}}$ for ${\sigma}\subset \{1,\cdots,r \}$ and $|\sigma|=s$ as the choice of dependent variables, then the free variables are $\eta^{\bar{\sigma}}$ for $\bar{\sigma}=\{1, \cdots\, r\} \setminus {\sigma}$. The associated basis series for this choice is
\begin{equation}\label{OldRule_1}
\begin{multlined}
I_{\bar{\sigma}}=\frac{1}{|\mathrm{det} \, B_{\sigma} |}\sum_{n_{\bar{\sigma}_{1}}, \cdots, n_{\bar{\sigma}_{r-s}}} \phi_{n_{\bar{\sigma}_{1}}, \cdots\, n_{\bar{\sigma}_{r-s}}} f(\eta^*)
                        \,\Gamma(-n_{{\sigma}_1}^{*})\dotsm\Gamma(-n_{{\sigma}_s}^{*})
                        \end{multlined}
                        \end{equation}
where ${n^{*}_i}$ are the solutions of ${n_i}$ for $i\in \sigma$, in terms of the free variables $\eta^{\bar{\sigma}}$, obtained by solving $B \eta +C=0$.
The column matrix $\eta^*$ is obtained from the matrix $\eta$ by substituting the solutions of dependent variables $\eta^{\sigma}$ while keeping the independent variables $\eta^{\bar{\sigma}}$ unchanged. The bracket equation can be rewritten to find the expression for $\eta^{*}_i$ as
\begin{align}\label{Bracket_matrix_solve}
& B \eta +C = B_{{\sigma}} \eta^{{\sigma}} + B_{\bar{\sigma}} \eta^{\bar{\sigma}} +C=0, \nonumber \\
& \eta^{{\sigma}} 
={\eta^{*}}^{\,\sigma}= -B^{-1}_{{\sigma}} \, B_{\bar{\sigma}} \eta^{\bar{\sigma}} - B^{-1}_{{\sigma}}\, C
\end{align}
Thus, we can rewrite Eq. \eqref{OldRule_1} as,
\begin{equation}\label{OldRule}
\begin{multlined}
I_{\bar{\sigma}}=\frac{1}{|\mathrm{det} \, B_{{\sigma}} |}\sum_{\eta^{\bar{\sigma}}} \phi_{\eta^{\bar{\sigma}}} f(\eta^{*})\,\Gamma(-{\eta^{*}}^{\,\sigma})
                        \end{multlined}
                        \end{equation}
where ${\eta^{*}}^{\,\sigma}=-B^{-1}_{{\sigma}} \, B_{\bar{\sigma}} \eta^{\bar{\sigma}} - B^{-1}_{{\sigma}}\, C$ and ${\eta^{*}}^{\,\bar{\sigma}}=\eta^{\bar{\sigma}}$.\\\\
 Similarly, for all other choices of free variables $\eta^{\bar{\sigma}}$ one will obtain a basis series of the form given in Eq.\eqref{OldRule}, unless the associated coefficient matrix $B_{\sigma}$, where $\sigma=\{1, \cdots\, , r\}\setminus \bar{\sigma}$, is singular.
                        \end{enumerate}
\indent If  the same series appears more than once in the final set of basis series then it should be taken only once. If divergent and/or null basis series appear then they should be discarded.
\item \textbf{Rule 5 :} This rule was formulated recently in \cite{gonzalez2016pochhammer} and states that if a Pochhammer symbol with negative index and negative integer base appears in the calculations, then it should be eliminated using the following limiting value of Pochhammer symbols for $k \in \mathbb{N}$ and $m \in \mathbb{N}$
\begin{equation}\label{Rule_5}
{
\lim_{\epsilon \to 0}-(k(m+\epsilon))_{-m-\epsilon}=(-k m)_{-m}=\frac{k}{k+1} \frac{(-1)^{m}(k m) !}{((k+1) m) !}}
\end{equation}
\end{itemize}
In conclusion, after applying the above set of rules, the MoB either yields a single term solution or a collection of basis series. In the latter case, the final solutions are obtained by finding the correct linear combinations of some of these basis series. To extract these combinations from the total set of basis series, one needs to find the convergence region of each basis series and find the largest subsets of these basis series having a nonempty intersection of their convergence regions. Each such subset gives one series solution of the integral by adding all the basis series in that subset.

\subsection{{The conic hull method for MB integrals}}\label{CHMss}
The CHMB, encoded in its very name, is based on the construction of a set of conic hulls,
associated with a given MB integral. Specific intersections of these conic hulls play the central role in determining the series representations of the integral. 

We begin by considering the general MB integral
\begin{align} \label{N_MB}
    I &(x_1,x_2,\cdots ,x_N) = \int\limits_{-i \infty}^{+i \infty} \frac{ d z_1}{2 \pi i} \cdots \int\limits_{-i \infty}^{+i \infty}\frac{ d z_N}{2 \pi i}\,\,  \frac{\prod\limits_{i=1}^{k} \Gamma^{a_i}({\bf e}_i\cdot{\bf z}+g_i)}{\prod\limits_{j=1}^{l} \Gamma^{b_j}({\bf f}_j\cdot{\bf z}+h_j)} x^{z_1}_{1} \cdots x^{z_N}_{N}
\end{align}
which is a $N$-fold integral where $a_i, b_j, k, l,$ and $N$ are positive integers with $k \geq N$ after possible cancellations due to the gamma functions of denominator and ${\bf e}_i$, ${\bf f}_j$ are $N$-dimensional coefficient vectors with ${\bf z}=(z_1, \cdots, z_N)$, while the variables $x_1 , \cdots , x_N$ can be complex. The integration contours have to be specified: in the following we consider the common situation where the set of poles of each gamma function of the numerator is not split in differents subsets by the contours.

The nature of the multivariable series solutions of Eq.\eqref{N_MB} is governed by the $N$-vector ${\bf\Delta}=\sum_{i=1}^{k} a_i\, {\bf e}_i - \sum_{j=1}^{l} b_j
\, {\bf f}_j$. Here, we restrict our discussion to the most frequently occurring case in the Feynman integral context, ${\bf\Delta}=0$, known as the degenerate case where several series representations of the MB integral will coexist, having different convergence regions. These are analytic continuations of each other provided that the scalar quantity $\alpha\doteq\text{Min}_{\vert\vert{\bf z}\vert\vert=1}(\sum_i a_i \vert{\bf e}_i\cdot{\bf y}\vert-b_j \sum_j\vert{\bf f}_j\cdot{\bf z}\vert)$ is positive. Further, we assume that we are in the nonresonant case where the poles of Eq.\eqref{N_MB} are of first-order only, thus avoiding in this short summary a discussion on the technical aspects of multivariate residues (we refer the reader to the Supplemental Material of \cite{ananthanarayan2020multiple} for a general and detailed discussion of the resonant case, as well as to \cite{Ananthanarayan:2020xpd} for a threefold resonant example).

In this scenario, the series solutions of Eq.\eqref{N_MB} that coexist can be obtained by the CHMB as follows (for later purpose, in the following we use slightly different notations than those of \cite{ananthanarayan2020multiple}):
\begin{itemize}
    \item Consider all possible $N$-combinations of the numerator gamma functions, denoting by the label $\sigma=(i_1, \cdots ,i_N)$ the combination containing the gamma functions $(\Gamma({\bf e}_{i_1}\cdot{\bf z}+g_{i_1}), \cdots, \Gamma({\bf e}_{i_N}\cdot{\bf z}+g_{i_N}))$.
    \item Not all combinations are retained, but only those for which the corresponding matrix 
      \begin{equation}    \label{MBmatrix}
A^{\sigma}=\left( \begin{array}{ccc}
A^{i_1}\\
A^{i_2}\\
.\\
A^{i_N}\\
\end{array}
 \right)
 \end{equation}
   is non-singular (we recall that $\sigma$ is not an index but a label),
where the $A^{i_j}, j\in\{1,...,N\}$, are $N$ particular rows, corresponding to the considered $N$-combination, of the $k\times N$ matrix
\begin{equation}
 A=\left( \begin{array}{ccc} \label{MBmatrix2}
\mathbf{e}_{1}\\
\mathbf{e}_{2}\\
.\\
\mathbf{e}_{k}\\
\end{array}
\right),
\end{equation}
where the $\mathbf{e}_{i}, i\in\{1,...,k\}$ are the coefficient vectors of Eq.(\ref{N_MB}) that are understood here as row matrices $\mathbf{e}_{i}= \left((\mathbf{e}_i)_1 \cdots (\mathbf{e}_i)_N \right)$.

\item For each retained $N$-combination with label $\sigma$, we associate a conic hull, with vertex at the origin and edges along the coefficient vectors ${\bf e}_{i_1}, \cdots , {\bf e}_{i_N}$, and we also associate a multiple series, called building block, which is the sum of the residues of only those poles emerging from the intersection of the singular hyperplanes of the gamma functions in the $N$-combination.
\item The series representations of the MB integral are finally obtained by finding all the largest subsets of conic hulls having a common intersection. Each of these largest subsets is associated with one distinct series representation of the MB integral which is obtained by adding the building blocks associated with the conic hulls that form this subset. 
\end{itemize}
Along with each series solution, the CHMB also provides in general, as a byproduct, a master series whose convergence region is conjectured to be either within or equal to that of the series solution. We refer the reader to \cite{ananthanarayan2020multiple,Ananthanarayan:2020xpd} for more technical details on the CHMB and now turn to discuss the MoB.

\section{Relation between the CHMB and the MoB} \label{Equivalence_Section}
As discussed in the previous section, both methods provide the final solution in terms of series, although the starting points of the methods are different. In the context of Feynman integrals computation, the CHMB starts with the MB representation of the Feynman integral, whereas, the MoB is applied to the Feynman parametric representation. We emphasize on the starting point because considering an appropriate starting MB representation for the CHMB will help us to readily find a connection with the MoB in the simplest situation of the nonresonant cases.

A given Feynman integral can have different but equivalent MB representations. Several techniques exist to derive these MB representations, one such being the Modified Method of Brackets (MMoB) \cite{Prausa:2017frh} which was motivated by MoB but differs from the latter by the fact that it uses the Inverse Mellin Transform instead of Ramanujan's Master Theorem. In the rest of this paper, we use the MMoB to derive MB representations as it facilitates to show, from a notational point of view, the connection between the CHMB and the MoB.

Using the MoB, suppose that we obtain, for a given integral, a bracket series of similar form as in Eq.\eqref{Bracket_Def}:
\begin{equation}\label{Bracket_Series_Comp}
 I=\sum_{\eta} \phi_{\eta}\,\,f(\eta)
                            \langle B \eta +C \rangle
                        \end{equation}
Then for the same integral one can derive the following MB bracket representation using the MMoB, 
\begin{equation}\label{MB_Rep_Bracket}
I= \prod_{{i=1,\dotsc,r}} \,\, \left( \int \frac{dz_i}{2 \pi i} \right)\,\,f(z_1,\dotsc,z_r)
                            \Gamma(-z_1)\cdots\Gamma(-z_r)\langle B Z +C \rangle
                        \end{equation}
where, we use $Z$ to denote the row matrix $Z= ( z_1, \cdots\, , z_r)^{T}$.\\
One can cross-check that Eq.\eqref{MB_Rep_Bracket} is correct by rewriting the explicit gamma functions in the above expression in terms of brackets, using
\begin{equation}
    \Gamma(-z_i)=\sum_{n_i=0}^{\infty} \phi_{n_i} \left< n_i-z_i \right>
\end{equation}
and subsequently removing all the MB integrals using the identity
\begin{equation}
\int \frac{d z_i}{2 \pi i} f(z_i) \left< n_i-z_1 \right>= f(n_i)
\end{equation}
This alternative approach was employed in \cite{Gonzalez:2021vqh} to show how some one-fold MB integrals can be computed from the MoB.

Now, from Eq.\eqref{MB_Rep_Bracket} one can obtain the following MB representation on applying Rule D of \cite{Prausa:2017frh} and taking $Z^{\bar{\sigma}_{1s}}$ for $\bar{\sigma}_{1s}=\{1, \cdots, r\}\setminus {\sigma}_{1s}=\{ {s+1}, \dotsc, r \}$ as the final integration variables :
\begin{equation}\label{MB_Rep_Comp_1}
I=\frac{1}{|\text{det} \, B_{{\sigma}_{1s}}|} \prod_{i \in\,\bar{\sigma}_{1s}} \,\, \left( \int \frac{dz_i}{2 \pi i} \right)\,\,f(z^*_1,\dotsc,z^*_s,z_{s+1},\dotsc,z_r)
                            \Gamma(-z^*_1)\cdots\Gamma(-z^*_s)\Gamma(-z_{s+1})\cdots\Gamma(-z_r)
                        \end{equation}
where $z^*_i$ are the solutions of $z_i$, for $i \in \sigma_{1s}$, in terms of final integration variables $Z^{\bar{\sigma}_{1s}}$  derived from the equation $B Z + C=0$. This equation was solved in Eq.\eqref{Bracket_matrix_solve}, thus $Z^{{\sigma}_{1s}}=-B^{-1}_{{\sigma}_{1s}} \, B_{\bar{\sigma}_{1s}} Z^{\bar{\sigma}_{1s}} - B^{-1}_{{\sigma}_{1s}}\, C$ and so Eq.\eqref{MB_Rep_Comp_1} becomes
\begin{equation}\label{MB_Rep_Comp_2}
I=\frac{1}{|\text{det} \, B_{{\sigma}_{1s}}|} \prod_{i \in \, \bar{\sigma}_{1s}} \,\, \left( \int \frac{dz_i}{2 \pi i} \right)\,\,f(-B^{-1}_{{\sigma}_{1s}} \, B_{\bar{\sigma}_{1s}} Z^{\bar{\sigma}_{1s}} - B^{-1}_{{\sigma}_{1s}}\, C,\,Z^{\bar{\sigma}_{1s}})
                            \Gamma(B^{-1}_{{\sigma}_{1s}} \, B_{\bar{\sigma}_{1s}} Z^{\bar{\sigma}_{1s}} + B^{-1}_{{\sigma}_{1s}}\, C)\Gamma(-Z^{\bar{\sigma}_{1s}})
                        \end{equation}
Using 
\begin{equation}\label{A_matrix}
A=\left(
\begin{array}{ccccc}
 B^{-1}_{{\sigma}_{1s}}\,B_{\bar{\sigma}_{1s}} \\\\
 -I_{r-s} \\
\end{array}
\right), \hspace{0.5cm}
C'=\left(
\begin{array}{ccccc}
B^{-1}_{{\sigma}_{1s}}\,C\\
 0_{r-s}\\
\end{array}
\right)
 \end{equation}
where $I_{r-s}$ denotes the identical matrix of order $r-s$, $C$ is the constant matrix defined in Eq.\eqref{B_Matrix} and $0_{r-s}$ denotes the null column matrix of length $r-s$,
 we get
\begin{equation}\label{MB_Rep_Comp}
I=\frac{1}{|\text{det} \, B_{{\sigma}_{1s}}|} \prod_{i \in \, \bar{\sigma}_{1s}} \,\, \left( \int \frac{dz_i}{2 \pi i} \right)\,\,f(-A \, Z^{\bar{\sigma}_{1s}}-C')
                            \Gamma(A \, Z^{\bar{\sigma}_{1s}}+C')
                        \end{equation}
where, as described before, for a column matrix $U$ with rows $U^i$ and length $l$,  we have $f(U)=f(U^1, \cdots, U^{l})$ and $\Gamma(U)=\prod_i \Gamma(U^i)$.

We note here that, at least in the context of Feynman integrals, the rational function $f$ does not contain gamma functions in its numerator whose arguments depend on the integration variables. Therefore, the last factor $ \Gamma(A \, Z^{\bar{\sigma}_{1s}}+C')$ contains all the ``numerator" gamma functions in Eq.\eqref{MB_Rep_Comp} which encapsulate the singular structure of the integrand. This is why we have used the matrix notation $A$  in Eq.(\ref{A_matrix}) as, once we apply the CHMB on the MB integral in Eq.(\ref{MB_Rep_Comp}), it will in fact correspond to the $A$ matrix of Eq.(\ref{MBmatrix2}).
The connection between the two methods can now be easily established by analysing Eqs.\eqref{Bracket_Series_Comp} and \eqref{MB_Rep_Comp}. Before that, let us assume that all poles of Eq.\eqref{MB_Rep_Comp} are of first-order, which corresponds to the nonresonant case in the context of the CHMB. Then, there is a maximum of $\binom{r}{r-s}=\binom{r}{s}$ possible building blocks for Eq.\eqref{MB_Rep_Comp}, each associated with one combination of $r-s$ numerator gamma functions out of the total number $r$, provided that their coefficient matrix is non-singular. Similarly, for Eq.\eqref{Bracket_Series_Comp}, we know by Rule 4 of the MoB that there is a maximum of $\binom{r}{s}$ possible basis series, corresponding to each possible choice of free summation variables.

A straightforward analysis, see below, shows that the set of all basis series of MoB is equal to the set of all building blocks in the CHMB. Indeed, one can point out a one-to-one mapping between a particular $N$-combination of numerator gamma functions in Eq.\eqref{MB_Rep_Comp} and a choice of free variables in Eq.\eqref{Bracket_Series_Comp}.

To see the correspondence, let us first consider the simple case where we take the $N$-combination of gamma functions to be $\bar{\sigma}_{1s}=\{s+1, \cdots, r \}$. We thus consider the poles of $(\Gamma(-z_{s+1}), \cdots, \Gamma(-z_{r}))$ which are at
$A^{\bar{\sigma}_{1s}}Z^{\bar{\sigma}_{1s}}+C'^{\,\bar{\sigma}_{1s}}=-I_{r-s}m^{\bar{\sigma}_{1s}}$, where $m=\{m_1, \cdots, m_r\}^{T}$ contains the summation variables such that $m_{i} \in \mathbb{N}$. From Eq.\eqref{A_matrix}, we have $A^{\bar{\sigma}_{1s}}=-I_{r-s}$ and $C'^{\,\bar{\sigma}_{1s}}=0_{r-s}$, so the poles are at $Z^{\bar{\sigma}_{1s}}=m^{\bar{\sigma}_{1s}}$. Hence,  summing over residues of all the poles leads to the corresponding building block
\begin{equation}
\begin{multlined}
\frac{1}{|\text{det}\, B_{{\sigma}_{1s}} |\,|\text{det} \, A^{\bar{\sigma}_{1s}} |}\sum_{m^{\bar{\sigma}}}\left(
\prod_{i\in\,\bar{\sigma}_{1s}} \frac{(-1)^{m_{i}}}{\Gamma(1+m_i)} \right) f(-A^{{\sigma}_{1s}} \, m^{\bar{\sigma}_{1s}}-C'^{\,{\sigma}_{1s}},\,m^{\bar{\sigma}_{1s}} )
                            \Gamma(A^{{\sigma}_{1s}} \, m^{\bar{\sigma}_{1s}}+C'^{\,{\sigma}_{1s}})
                        \end{multlined}
                        \end{equation}
We note that from Eq.\eqref{A_matrix}, $|\text{det}(A^{\bar{\sigma}_{1s}})|=|\text{det}(-I_{r-s})|=1$, $A^{{\sigma}_{1s}}=B^{-1}_{{\sigma}_{1s}} \, B_{\bar{\sigma}_{1s}}$, $C'^{\,{\sigma}_{1s}}=B^{-1}_{{\sigma}_{1s}}\, C$, hence the above expression is equal to
\begin{equation}
\begin{multlined}
\frac{1}{|\text{det}\, B_{{\sigma}_{1s}} |}\sum_{m^{\bar{\sigma}}}\left(
\prod_{i\in\,\bar{\sigma}_{1s}} \frac{(-1)^{m_{i}}}{\Gamma(1+m_i)} \right) f(-B^{-1}_{{\sigma}_{1s}} \, B_{\bar{\sigma}_{1s}} m^{\bar{\sigma}_{1s}} - B^{-1}_{{\sigma}_{1s}}\, C,\,m^{\bar{\sigma}_{1s}})
                            \Gamma(B^{-1}_{{\sigma}_{1s}} \, B_{\bar{\sigma}_{1s}} m^{\bar{\sigma}_{1s}} + B^{-1}_{{\sigma}_{1s}}\, C)
                        \end{multlined}
                        \end{equation}
The above series is the same as the basis series associated with the choice of free summation variables $\eta^{\bar{\sigma}_{1s}}$ for $\bar{\sigma}_{1s}=\{ {s+1}, \cdots, {i_{r}\}}$ derived using Eq.\eqref{OldRule}. Thus, for this simple case we see the correspondence between the building block of the $N$-combination with label $\bar{\sigma}_{1s}$ and the basis series with free variables $\eta^{\bar{\sigma}_{1s}}$.

Let us now extend this for an arbitrary choice $\bar{\sigma}=\{\bar{\sigma}_1, \cdots, \bar{\sigma}_{r-s}\}$ for $\bar{\sigma}\subset \{1,\cdots, r\}$, by showing that the building block associated with the $N$-combination of gamma functions $(\Gamma(-z_{\bar{\sigma}_1}), \cdots, \Gamma(-z_{\bar{\sigma}_{r-s}}))$ in Eq.\eqref{MB_Rep_Comp_1}, is the same as the basis series for the choice of free variables $\eta^{\bar{\sigma}}=(n_{\bar{\sigma}_1} \cdots \, n_{\bar{\sigma}_{r-s}})^{T}$ given in Eq.\eqref{OldRule}. We define $\sigma=\{1,\cdots,r\}\setminus\bar{\sigma}$, and assume $|\text{det} \, A^{\bar{\sigma}}|\neq 0$ and $|\text{det} \, B_{{\sigma}}|\neq 0$, so that both the associated building block and basis series exist.

To find the building block, we first need to find the poles originating from the intersection of singular hyperplanes of the combination of gamma functions $(\Gamma(-z_{\bar{\sigma}_1}), \cdots, \Gamma(-z_{\bar{\sigma}_{r-s}}))$  which, by Eq.\eqref{MB_Rep_Comp}, is equal to $\Gamma(A^{\bar{\sigma}}Z^{\bar{\sigma}_{1s}}+C'^{\,\bar{\sigma}})$. Then the poles can be found by solving $A^{\bar{\sigma}}Z^{\bar{\sigma}_{1s}}+C'^{\,\bar{\sigma}}=-I_{r-s}m^{\bar{\sigma}}$, which gives $Z^{\bar{\sigma}_{1s}}=-(A^{\bar{\sigma}})^{-1}m^{\bar{\sigma}}-(A^{\bar{\sigma}})^{-1}\,C'^{\,\bar{\sigma}}$. Summing over the residues of all these poles then leads to the building block
\begin{align}\label{Arbitrary_BB_1}
\frac{1}{|\text{det}\, B_{{\sigma}_{1s}} |\,|\text{det} \, A^{\bar{\sigma}} |}\sum_{m^{\bar{\sigma}}}\left(
\prod_{i\in\bar{\sigma}} \frac{(-1)^{m_{i}}}{\Gamma(1+m_i)} \right)  f(Z^*) \, \Gamma(-{Z^*}^{\,\sigma})
                        \end{align}
where $Z^*$ is a column matrix of length $r$ whose elements are, ${Z^*}^{\,\sigma}=A^{\sigma}(A^{\bar{\sigma}})^{-1}m^{\bar{\sigma}}-C'^{\, \sigma}+A^{\sigma}(A^{\bar{\sigma}})^{-1}\,C'^{\,\bar{\sigma}}$ and ${Z^*}^{\,\bar{\sigma}}=m^{\bar{\sigma}}$.
\\
We now substitute the identities (proved in Appendix B)
\begin{align}
&(B_{\sigma})^{-1} B_{\bar{\sigma}}+A^{\sigma}(A^{\bar{\sigma}})^{-1}=0 \label{Identity1}\\
 &(B_{\sigma})^{-1}\,C-C'^{\, \sigma} +A^{\sigma}(A^{\bar{\sigma}})^{-1}\,C'^{\,\bar{\sigma}}=0 \label{Identity2} \\
 &|\text{det}\, B_{{\sigma}_{1s}} |\,|\text{det} \, A^{\bar{\sigma}} |=|\text{det}\, B_{{\sigma}} |  \label{Identity3}
\end{align}
 in Eq.\eqref{Arbitrary_BB_1} to transform it as
\begin{align}\label{Arbitrary_BB}
\frac{1}{|\text{det}\, B_{{\sigma}}|}\sum_{m^{\bar{\sigma}}}\left(
\prod_{i\in\bar{\sigma}} \frac{(-1)^{m_{i}}}{\Gamma(1+m_i)} \right)  f(Z^*) \,\,\, \Gamma(-{Z^*}^{\,\sigma})
                        \end{align}
where, ${Z^*}^{\,\sigma}=-\,(B_{\sigma})^{-1} B_{\bar{\sigma}}m^{\bar{\sigma}}-(B_{\sigma})^{-1}\,C$ and ${Z^*}^{\,\bar{\sigma}}=m^{\bar{\sigma}}$.\\\\
The above series, which is the building block of the $N$-combination $\bar{\sigma}$ in Eq.\eqref{MB_Rep_Comp}, is the same as the basis series associated with the choice of free variables $\eta^{\bar{\sigma}}$ given in Eq.\eqref{OldRule}. As $\bar{\sigma}$ is an arbitrary choice, this proves the one-to-one correspondence between basis series and building blocks and establishes the equivalence between the two methods.\\ However, this equivalence stops here because, in the next crucial step of the calculations, the CHMB can extract, from the set of building blocks, the relevant linear combinations which form the series representations of the integral, whereas MoB cannot do a similar task from the set of its basis series without a complete convergence analysis of the basis series. We will discuss about this point and other limitations of MoB in the next section.
                        
                        We remark that the equivalence discussed above extends the work \cite{Suzuki:2003jn}, which discussed the equivalence between MB and NDIM at one-loop order, still (implictly) in nonresonant situations.
\section{{Merits of CHMB in comparison with MoB}}\label{Limitation_Section}
Based on the connection established in the previous section, we now use known properties of the MB approach to make a few remarks and discuss the limitations of the MoB.

As we have just seen, in the nonresonant case both methods provide the same set of basis series or building blocks. 
But this is not true in the resonant case where some termwise divergent basis series appear in the MoB, which are then discarded by Rule 4, leading to incomplete solutions even after proper convergence analysis of the retained convergent basis series. This does not occur with the CHMB. To explain this, let us have a closer look at what happens in the resonant case. In the context of MB integrals, the resonant case is defined as the case for which the MB integrand has poles of order greater than one. Hence, as shown in \cite{ananthanarayan2020multiple}, when the MB integrals has several folds this requires nontrivial computations using the machinery of multivariate residues, in order to derive the corresponding series. Now, from the connection established in Section \ref{Equivalence_Section}, we see that by applying Eq.\eqref{OldRule} to derive the basis series, the MoB is doing as if, in the context of MB, one would treat all poles as first-order poles and thus, whenever higher-order poles appear (in MB), MoB correspondingly gives a term-wise divergent basis series. However, for these poles, the CHMB gives a convergent series, by a proper singularity analysis using multivariate residues theory. Thus, the rule discarding the divergent basis series in the MoB (Rule 4) leads to an incomplete result for those series representations whose convergence region intersects with the convergence regions of series obtained from the higher-order poles by the CHMB. This is explicitly shown in an example in Section \ref{resonantExample}.

Next, let us discuss the ad-hoc Rule 5 of Section \ref{MoB_Rules}, which was introduced in \cite{gonzalez2016pochhammer} to solve another limitation of MoB. Exploiting the connection presented in Section \ref{Equivalence_Section} we have shown in Appendix A that Rule 5 naturally emerges from the residue calculation in the CHMB. Therefore, there is no need to distinguish these cases in the CHMB: they are directly treated in a correct way.

Finally, we end this section with a discussion on the limitation of the MoB coming from the fact that it hinges on the analysis of the convergence regions of the basis series, in order to obtain their grouping into the final series solutions. Indeed, in Section \ref{MoB_Rules}, we have noted that, for $r>s$, one gets a set of basis series whose correct linear combinations give the series representations of the integral. However, the derivation of these linear combinations requires the knowledge of the convergence region of each of the basis series, which is a significant drawback of the MoB. Indeed, these basis series are primarily of the hypergeometric type, whose convergence regions can be difficult to find. For instance, when the series have three or more variables, their convergence studies are often open problems in mathematics (to have an idea of the complexity of such problems, we refer the reader to \cite{srivastava1985multiple} for a list of the convergence regions of the 205 triple gaussian hypergeometric series of lowest order). Moreover, the number of such basis series can be huge and, as already emphasized, one has to determine the convergence region of each of them before being able to derive series representations from them. As an example, let us consider the case of the hexagon conformal Feynman integral studied in \cite{PhysRevD.101.066006,Ananthanarayan:2020ncn}. It was shown in \cite{PhysRevD.101.066006} and confirmed in \cite{Ananthanarayan:2020ncn} that there are not less than 2530 building blocks (i.e. basis series) of nine variables, although the series representation obtained from CHMB and given in \cite{Ananthanarayan:2020ncn} for the hexagon is built from only 26 of these series. It is obvious that MoB could not solve a problem of such a complexity. On the other hand, the same problem is bypassed in the CHMB, where the groupings of the series can be achieved by only considering the intersections of conic hulls associated with each building block.
We note here that one could adapt the conic hull technique to the MoB, by first solving the linear equations for all particular choices $\sigma$ of the free variables (with non-singular associated $A^\sigma$ matrix in Eq.(\ref{MBmatrix})), and then by building conic hulls from the coefficients of the independant variables, the intersections of these conic hulls giving the series solutions, as in CHMB. However, we do not see the usefulness of such an improvement of the MoB, since it would then become redundant\footnote{Only for nonresonant cases: resonant cases would still not be properly solved by MoB.} with the CHMB and would not give any advantage over the latter, since we have shown that any case solved by MoB can be solved by CHMB. Moreover, as we have just seen above, MoB suffers from other drawbacks which would not be solved by these considerations. At last, the CHMB has already been fully automatized in the \texttt{MBConicHulls.wl} \textit{Mathematica} package \cite{ananthanarayan2020multiple}, at a very user-friendly level and, therefore, can be easily used as such without the need of a clone technique.

\section{An Example}\label{Box_Diagonal_Solve}
Let us now illustrate the discussion of the preceding section by the evaluation of a two-loop scalar dimensionally regularized Feynman integral, using both the MoB and the CHMB. 
We will first consider the (nonresonant) generic propagator powers case and, in a second part, the (resonant) case of unit propagator powers. The results have been numerically verified for both cases, by setting $\epsilon=2-D/2$ to a small value using FIESTA and also by direct integration of Eq.\eqref{Box_MB_General}. This also helps, in the resonant case, to confirm the regions for which the MoB leads to an incorrect result.

\begin{figure}[ht]
\centering
\includegraphics[width=10cm]{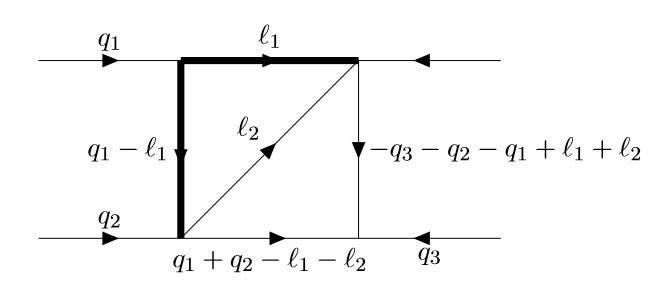}
\caption{The 2-Loop box diagonal integral with two massive and three massless lines.}
\end{figure}
            
The corresponding Feynman integral in Euclidean space-time is:
\begin{equation}
\begin{multlined}
I(a_1,\cdots,a_5)=\int  \frac{\mathrm{d}^D l_1}{{\mathrm{\pi }^{D/2}}}\int  \frac{\mathrm{d}^D l_2}{{\mathrm{\pi }^{D/2}}} \frac{1}{[l_1^2+m^2]^{a_1}[l_2^2]^{a_2}[(q_1-l_1)^2+m^2]^{a_3}[(q_1+q_2-l_1-l_2)^2]^{a_4}}\\
\times \frac{1}{[(-q_3-q_2-q_1+l_1+l_2)^2]^{a_5}}
\label{eq:Feynman_Integral_5}
\end{multlined}
\end{equation}
Under the kinematic constraints
\begin{equation}\label{Kinematic_2}
q_1^2=0; \,\quad q_2^2=0; \,\quad q_3^2=0; \, \quad q_1\cdot q_2=\frac{s}{2};\, \quad q_2\cdot q_3=\frac{t}{2};\, \quad q_1\cdot q_3=-\frac{s}{2}-\frac{t}{2};
\end{equation}
we get the following Schwinger parametrized form of the Feynman integral:
\begin{equation}
\displaystyle{I(a_1,\cdots,a_5)=\frac{1}{\Gamma(a_1)\cdots\Gamma(a_5)}\int\limits_0^\infty\mathrm{dx_1}\,{x_1}^{a_1-1}\cdots\int\limits_0^\infty \mathrm{dx_5}\, {x_5}^{a_5-1} \frac{e^{-\frac{F}{U}-(x_1+x_3)m^2}}{U^{D/2}}}
\label{eq:Schwinger_2}
\end{equation}
where, $F=sx_1x_2x_4+tx_2x_3x_5$ and $U=(x_1+x_2+x_3)(x_2+x_4+x_5)-(x_2)^2$.

Next, applying Rules 1, 2 and 3 of the MoB to Eq.\eqref{eq:Schwinger_2} gives us the following bracket series:
\begin{equation}\label{Box_Bracket_1}
\begin{multlined}
I(a_1,\cdots,a_5)=\sum_{n_1,\dots,n_9,n_a}\phi_{n_1,\cdots,n_9,n_a}{(m^2)^{n_1}(s)}^{n_2}(t)^{n_3}\frac{\left\langle n_{23456}+D/2 \right\rangle\left\langle n_{78}-n_{145} \right\rangle\left\langle n_{9a}-n_{56} \right\rangle}{\Gamma(a_1)\cdots\Gamma(a_5)\Gamma(n_{23}+D/2)}\\
\times \frac{ \left\langle n_{27}+a_1 \right\rangle\left\langle n_{2346}+a_2 \right\rangle\left\langle n_{38}+a_3\right\rangle \left\langle n_{29}+a_4 \right\rangle\left\langle n_{3a}+a_5 \right\rangle}{\Gamma(-n_{145})\Gamma(-n_{56})}
\end{multlined}
\end{equation}
where $n_a$ denotes $n_{10}$. Here, and in the rest of this section we use the notation $x_{123} =x_1+x_2+x_3$.

Similarly, using the rules of the MMoB we obtain the following MB bracket representation,
\begin{equation}\label{Box_MB_Bracket_1}
\begin{multlined}
I(a_1,\cdots,a_5)= \prod_{i=1}^{10} \left( \int \frac{d z_i}{2 \pi i} \Gamma(-z_i) \right) {(m^2)^{z_1}(s)}^{z_2}(t)^{z_3}\frac{\left\langle z_{23456}+D/2 \right\rangle\left\langle z_{78}-z_{145} \right\rangle\left\langle z_{9a}-z_{56} \right\rangle}{\Gamma(a_1)\cdots\Gamma(a_5)\Gamma(z_{23}+D/2)}\\
\times \frac{ \left\langle z_{27}+a_1 \right\rangle\left\langle z_{2346}+a_2 \right\rangle\left\langle z_{38}+a_3\right\rangle \left\langle z_{29}+a_4 \right\rangle\left\langle z_{3a}+a_5 \right\rangle}{\Gamma(-z_{145})\Gamma(-z_{56})}
\end{multlined}
\end{equation}
where $z_a$ is used to denote $z_{10}$. 

As there are 10 integration variables and 8 brackets, the final MB integral will be two-fold. All possible choices of final integration variables for which the system of linear equations obtained from the vanishing of the bracket equations in Eq.\eqref{Box_MB_Bracket_1} are solvable lead to equivalent MB representations. Hence, without loss of generality, let us take $z_1$ and $z_2$ as the final integration variables to get
\begin{equation}\label{Box_MB_General}
\begin{multlined}
I(a_1,\cdots,a_5)=  \frac{1}{\Gamma(a_1)\cdots\Gamma(a_5)} \int \frac{d z_1}{2 \pi i} \int \frac{d z_2}{2 \pi i} \Gamma(-z_1)\Gamma(-z_2) \Gamma(-z^*_3) \cdots \Gamma(-z^*_{10})  \\
\times \frac{(m^2)^{z_1}(s)^{z_2}(t)^{z^*_3}}{\Gamma(z_2+z^*_3+D/2)\Gamma(-z_1-z^*_4-z^*_5)\Gamma(-z^*_5-z^*_6)}
\end{multlined}
\end{equation}
where $z^*_i$, for $ 3\leq i \leq 10$ are solutions of bracket equations in Eq.\eqref{Box_MB_Bracket_1} in terms of $z_1$ and $z_2$.
\begin{align} \label{Dependent_Sub}
    & z^*_3 = D-a_{12345}-z_{12}, && z^*_4 = a_{45}-D/2, && z^*_5 = a_2-D/2, && z^*_6 = a_{13}-D/2+z_1\nonumber \\
    & z^*_7 = -a_1-z_2, && z^*_8 = a_{1245}-D+z_{12}, && z^*_9 = -a_4-z_2, && z^*_{10} = a_{1234}-D+z_{12}
\end{align}
Note that ${\bm\Delta}=(0,0)$ \cite{ananthanarayan2020multiple} for the MB representation in Eq.\eqref{Box_MB_General}. Hence, it is a degenerate situation where several convergent series representations will coexist.
\subsection{Nonresonant case}
To begin with, let us consider the simple case of generic propagator powers, which corresponds to a nonresonant situation. This will allow us to explicitly show that both methods yield the same set of basis series (or building blocks) and to illustrate the one-to-one correspondence pointed out in Section \ref{Equivalence_Section}.

We have to apply the MoB and the CHMB to Eq.\eqref{Box_Bracket_1} and Eq.\eqref{Box_MB_General}, respectively. First, note that in Eq.\eqref{Box_Bracket_1} we have a 10-fold summation and eight brackets, thus there are $\binom{10}{8}=45$ ways of choosing two free variables, hence a maximum of 45 possible basis series. This is the same as the maximum number of possible building blocks for Eq.\eqref{Box_MB_General}, as there are ten gamma functions in the numerator of the MB integrand and we have to choose two at a time.

Next, in 24 of the 45 possible choices of two free variables, the corresponding coefficient matrix $B$ of Rule 4 is found to be singular, thus we will have only 21 choices of free variables giving basis series in the MoB. In the CHMB picture, this also means that only 21 choices of 2-combinations will have an associated two-dimensional conic hull. Indeed, take, for example, the 2-combination $(\Gamma(-z_1),\Gamma(-z^*_6))$, whose conic hull is one-dimensional as the coefficient vectors ${\bf e}_1=(-1,0)$ and ${\bf e}_6=(1,0)$ are parallel. This 2-combination does not give birth to a building block. Therefore, by the connection established in Section \ref{Equivalence_Section}, the choice of free variable $(n_1,n_6)$ in the MoB will also not lead to any basis series as its corresponding matrix $B$ is singular. In fact, one can readily remark that all choices of free variables, in the MoB, which include either $n_4$ or $n_5$, will give no basis series as the corresponding gamma functions $\Gamma(-z^*_4)$ and $\Gamma(-z^*_5)$, of Eq.\eqref{Box_MB_General}, have no poles.

Among the 21 choices for which we get non-zero series, let us now evaluate the basis series for the choice of free variable $(n_1,n_7)$ using the MoB and also the building block associated with $(\Gamma(-z_1),\Gamma(-z^*_7))$ in the CHMB, in order to show that both are equal, as claimed in Section \ref{Equivalence_Section}.

\begin{itemize}
    \item basis series for the choice of free variable $(n_1,n_7)$ using the MoB:
    
    Solving the brackets in Eq.\eqref{Box_Bracket_1} by setting them to zero leads to the following
    solutions
    \begin{align} \label{Dependent_Sub_MoB}
    & n^*_2 = -a_1-n_7, && n^*_3 = -a_{2345}+D-n_1+n_7, && n^*_4 = a_{45}-D/2\nonumber \\
    &  n^*_5 = a_2-D/2, && n^*_6 = a_{13}-D/2+n_1, && n^*_8 = a_{245}-D+n_1-n_7
    \nonumber \\
    & n^*_9 = a_1-a_4+n_7, && n^*_{10} = a_{234}-D+n_1-n_7
\end{align}
    
    Substituting these values in Eq.\eqref{OldRule} gives the basis series,
    \begin{align} \label{Series_13}
   & S_{1,7}  = \frac{\Gamma (2-\epsilon-{a_2}) \Gamma (2-\epsilon-a_{45}) (s)^{-a_1} (t)^{4-a_{2345}-2\epsilon}}{\Gamma ({a_1}) \cdots \Gamma ({a_5})} \sum_{n_1,n_7=0}^{\infty} \frac{ \Gamma \left(-a_1+a_4-n_7\right) }{ \Gamma \left(-2 \epsilon -a_{245}-n_1+4\right) } 
 \nonumber \\ & \times \frac{ \Gamma \left(2-\epsilon -a_{13}-n_1\right)\Gamma \left(2 \epsilon+a_{2345}+n_1-n_7-4\right) \Gamma \left(4-2 \epsilon-a_{234}-n_1+n_7\right) }{ \Gamma \left(-2 \epsilon -a_{123}-n_1+4\right) \Gamma \left(-3 \epsilon -a_{12345}-n_1+6\right)}
    \\ \nonumber & \times    
   \Gamma \left(a_1+n_7\right)  \Gamma \left(-2 \epsilon
   -a_{245}-n_1+n_7+4\right) \frac{1}{n_1! \, n_7!}\left(-\frac{m^2}{t}\right)^{n_1}\left(-\frac{t}{s}\right)^{n_7}
    \end{align}
    where we have also substituted $D=4-2\epsilon$.
\item Building block for the choice of 2-combination $(\Gamma(-z_1),\Gamma(-z^*_7))$ using the CHMB: 

The poles associated with this 2-combination are at $(z_1,z_2)=(m_1,-m_2-a_1)$ for $m_i \geq 0 $. Following \cite{ananthanarayan2020multiple}, one shifts the poles to the origin by performing the change of variables $z_1 \to z_1+m_1$ and $z_2 \to z_2-m_2-a_1$ on Eq.\eqref{Box_MB_General} to get, for the MB integrand,

\begin{align}
& \frac{\Gamma (D/2-{a_2}) \Gamma (D/2-a_{45}) (t)^{D-a_{12345}} }{\Gamma ({a_1}) \cdots \Gamma ({a_5})} \left( \frac{m^2}{t} \right)^{m_1+z_1} \left( \frac{s}{t} \right)^{-m_2-a_1+z_2} \Gamma(a_1+m_2-z_2) \\ \nonumber & \times \frac{\Gamma(-m_1-z_1)\Gamma(-m_2+z_2)\Gamma(a_{2345}-D+m_1-m_2+z_{12})\Gamma(D/2-a_{13}-m_1-z_1)  }{\Gamma(3D/2-a_{12345}-m_1-z_1)\Gamma(D-a_{245}-m_1-z_1)\Gamma(D-a_{123}-m_1-z_1)} \\ \nonumber & \times \Gamma(D-a_{245}-m_1+m_2-z_{12})\Gamma(a_4-a_1-m_2+z_2)\Gamma(D-a_{234}-m_1+m_2-z_{12})
\end{align}
Noting that only $\Gamma(-m_1-z_1)$ and $\Gamma(-m_2+z_2)$ are singular at the origin, one can extract out the singularity explicitly by applying the generalized reflection formula $\Gamma(z-m)=\frac{(-1)^m \Gamma(1+z)\Gamma(1-z)}{z \, \Gamma(m+1-z)}$ for $m \in \mathbb{Z}$. Then, making use of the residue theorem to find the explicit series, the latter turns out to be the same as in Eq.\eqref{Series_13}.
\end{itemize}

Hence, we see that the one-to-one correspondence proposed in Section \ref{Equivalence_Section} holds for the above case. A straightforward but lengthy calculation mimicking the above steps confirms that this is also true for all the remaining 20 choices.

\subsubsection{Solutions and Convergent Regions}
\begin{figure}[ht]
\centering
\includegraphics[width=8cm]{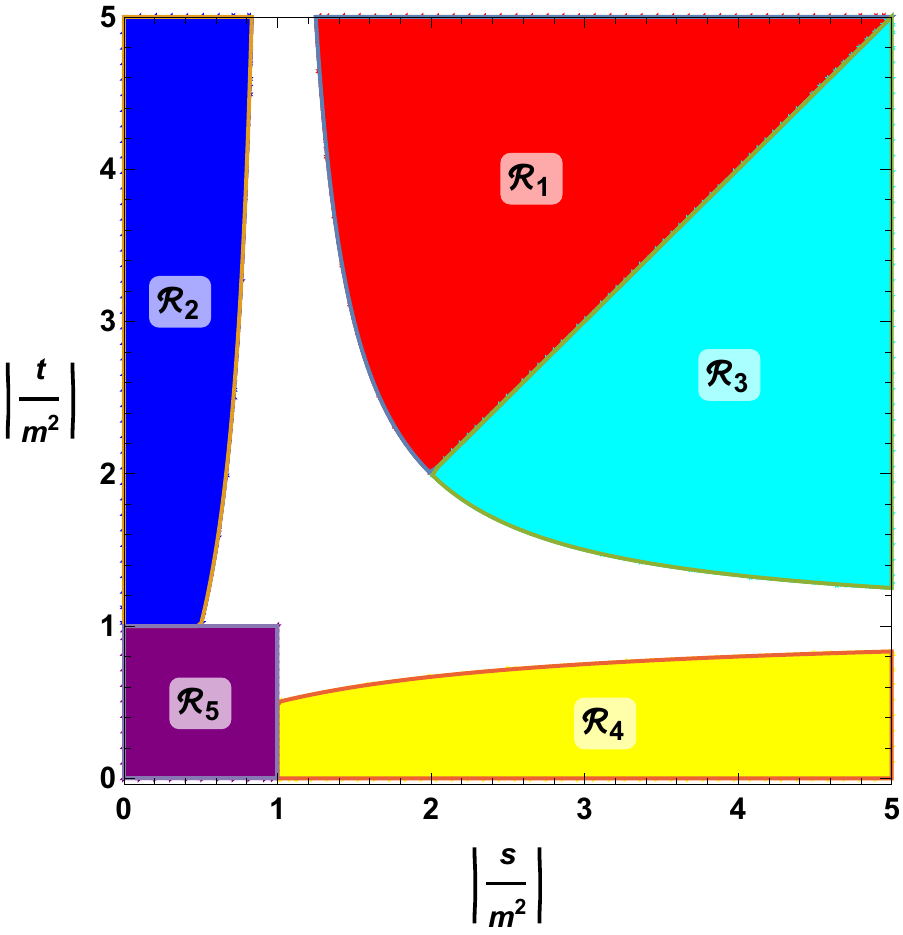}
\caption{Region of convergence of the series representations of 2-Loop box diagonal integral.}\label{PlotROC}
\end{figure}

The correct groupings of the 21 basis series or building blocks yield the actual series solutions of the Feynman integral in Eq.\eqref{eq:Feynman_Integral_5}. This can be achieved either by finding the individual convergence regions of each basis series and studying their intersections, or by using the CHMB, which relies only on the geometry of the conic hulls associated with each basis series. As the basis series involved here are simple double hypergeometric series, the convergence regions can be easily found and, therefore, the groupings as well. This can be done using Horn's theorem or even the simpler approach resting on the fact that the convergence region of a multiple hypergeometric series is independent of the choice of its parameters. Indeed, from the latter approach it is straighforward to conclude that the series representations $S_5$, $S_2$ and $S_4$ (see the \textit{Mathematica} notebook BoxDiagonal.nb given as an ancillary file for the explicit formulas of the $S_i$) of the two-loop box diagonal Feynman integral with two massive lines converge as the Appell $F_3$ hypergeometric series and its basic analytic continuations (see Eq.(67) and its symmetrical copy in Chapter 9 of \cite{srivastava1985multiple}). Alternatively, we can also use the package \texttt{MBConicHulls.wl} to find the groupings. Both procedures identically lead to five different solutions, which we list below, according to their regions of convergence (see Fig. \ref{PlotROC} for a plot of the latter):
\begin{itemize}
    \item Region $\bm{\mathcal{R}_{1}}=$\Big\{ $\Big| \dfrac{m^2}{s} \Big|<1 \,\,\cap\,\, \Big|\dfrac{s}{t}\Big|<1 \,\,\cap\,\, \Big|\dfrac{m^2}{s}\Big|+\Big|\dfrac{m^2}{t}\Big|<1 $ \Big\}. Here the basis series that have this convergence region in common are associated with the free variables $(n_1,n_2)$, $(n_2,n_6)$, \bm{$(n_1,n_8)$}, $(n_6,n_8)$, $(n_7,n_8)$, $(n_1,n_a)$, $(n_6,n_a)$, $(n_7,n_a)$, $(n_8,n_9)$ and $(n_9,n_a)$.
    \item Region $\bm{\mathcal{R}_{2}}=$\Big\{ $\Big| \dfrac{s}{m^2} \Big|<1 \,\,\cap\,\, \Big|\dfrac{m^2}{t}\Big|<1 \,\,\cap\,\, \Big|\dfrac{s}{t}\Big|+\Big|\dfrac{s}{m^2}\Big|<1 $ \Big\}. In this case, the basis series that share this convergence region are associated with the free variables $(n_1,n_2)$, $(n_2,n_6)$, \bm{$(n_2,n_8)$} and $(n_2,n_a)$.
    \item Region $\bm{\mathcal{R}_{3}}=$\Big\{ $\Big| \dfrac{t}{s} \Big|<1 \,\,\cap\,\, \Big|\dfrac{m^2}{t}\Big|<1 \,\,\cap\,\, \Big|\dfrac{m^2}{s}\Big|+\Big|\dfrac{m^2}{t}\Big|<1 $ \Big\}. The basis series  that share this convergence region are associated with the free variables $(n_1,n_3)$, $(n_3,n_6)$, \bm{$(n_1,n_7)$}, $(n_6,n_7)$, $(n_7,n_8)$, $(n_1,n_9)$, $(n_6,n_9)$, $(n_7,n_a)$, $(n_8,n_9)$ and $(n_9,n_a)$.
    \item Region $\bm{\mathcal{R}_{4}}=$\Big\{ $\Big| \dfrac{m^2}{s} \Big|<1 \,\,\cap\,\, \Big|\dfrac{t}{m^2}\Big|<1 \,\,\cap\,\, \Big|\dfrac{t}{s}\Big|+\Big|\dfrac{t}{m^2}\Big|<1 $ \Big\}. Here the basis series are those associated with the free variables $(n_1,n_3)$, $(n_3,n_6)$, \bm{$(n_3,n_7)$} and $(n_3,n_9)$.
    \item Region $\bm{\mathcal{R}_{5}}=$\Big\{
 $ \Big|\dfrac{s}{m^2} \Big|<1 \,\,\cap\,\, \Big|\dfrac{t}{m^2}\Big|<1 \,\,\cap\,\, \Big|\dfrac{m^2}{s}\Big|+\Big|\dfrac{m^2}{t}\Big|>1$\Big\}. In this last case, there is only one series, which comes from the free variable \bm{$(n_2,n_3)$}.
\end{itemize}
The free-variable in bold indicates that the convergence region of its associated basis series is identical to the convergence region of the series solution to which it belongs. This concept comes from the important notion of master series in the CHMB, which is discussed in detail in \cite{ananthanarayan2020multiple}. For higher-dimensional MB integrals, master series can be very useful for numerical checks, as was shown in \cite{Ananthanarayan:2020ncn}, and also to simplify the convergence analysis.

Let us summarize what we have done in this section. We have illustrated on one explicit example that MoB and MB are giving equivalent sets of basis series (building blocks) in non-resonant cases, but that the series representations built from subsets of these basis series cannot be derived from MoB without a convergence analysis of all the basis series, whereas CHMB does not require such a study. As, in general, the convergence regions of these basis series will be too difficult to find, one concludes that the series representations cannot be obtained from MoB in more complicated cases.
\subsection{Resonant Case and Numerical comparison\label{resonantExample}}
Let us now move to the case where all the powers of the propagators are set to one. This is a resonant case in the context of the CHMB as poles of higher order appear in Eq.\eqref{Box_MB_General}. Correspondingly, one can see that this is manifested as an emergence of some divergent basis series in the MoB. One can readily see this fact in Eq.\eqref{Series_13} by substituting $a_i=1$ for $1\leq i \leq 5$. This indeed leads $S_{1,7}$ to become a termwise diverging series, which is then discarded by Rule 4 of MoB. The same is not true for the CHMB as it can properly handle resonant cases. Indeed, the contribution due to the corresponding 2-combination $(\Gamma(-z_1),\Gamma(-z^*_7))$ derived earlier, is  still a converging series, but now logarithmic and of the form

\begin{align} \label{Series_13_Resonant}
   &\Gamma (1-\epsilon) \Gamma (-\epsilon) (s)^{-1} (t)^{-2\epsilon} \sum_{m_1,m_2=0}^{\infty} \frac{ 
    \Gamma(2\epsilon+m_1-m_2)\Gamma^2(1-2\epsilon-m_1+m_2)}{ \Gamma(1-3\epsilon-m_1)\Gamma^2(1-2\epsilon-m_1) } 
  \\ \nonumber & \times \Gamma(-\epsilon-m_1) \left( \text{log}\left[\frac{s}{t}\right]+\psi(2\epsilon+m_1-m_2) + \psi(1+m_2) - 2 \psi(1-2\epsilon-m_1+m_2) \right)
    \\ \nonumber & \times  \frac{1}{m_1! \, m_2!}\left(-\frac{m^2}{t}\right)^{m_1}\left(\frac{t}{s}\right)^{m_2}
    \end{align}
    
    which is found either by hand or by using the \texttt{MBConicHulls.wl} package.

In the MoB, several choices of free variables other than $(n_1,n_7)$ also lead to divergent basis series, discarded by Rule 4. Due to which the MoB yields, even after proper convergence analysis, incomplete results for those series solutions where logarithmic contributions, like Eq.(\ref{Series_13_Resonant}), emerge in CHMB.

We have verified the solutions from both the methods numerically with direct integration of the MB integral in Eq.\eqref{Box_MB_General} and with FIESTA \cite{smirnov2016fiesta4}. For the nonresonant case, a good numerical matching was obtained with the results of both methods (which were identical). In the resonant case only the results from CHMB had a perfect agreement. These results do not match (except in $\bm{\mathcal{R}_{5}}$) with the solutions coming from the MoB because the regions $\bm{\mathcal{R}_{i}}$, $1\leq i \leq 4$ have at least one choice of free variable which, in the nonresonant case, gives a convergent basis series but which is transformed into a divergent basis series in the resonant case, like the $(n_1,n_7)$ case shown above. The detailed numerical analysis is presented in the BoxDiagonal.nb \textit{Mathematica} notebook.

\section{Conclusion and Discussion}\label{Conclusion_Section}
Of the many methods available today for the evaluation of Feynman integrals, one approach is the MoB. This rests on Ramanujan's Master Theorem, and is not limited to Feynman integrals, and is also helpful in evaluating integrals arising in other branches of physics and mathematics. Nevertheless, some of its foundational issues have not been resolved in their entirety: although successful in some instances, there are important classes of examples for which one can show that MoB has not been settled satisfactorily.

A very recent systematic analytic method to evaluate multiple MB integrals has been developed, based on a simple geometric analysis using conic hulls, namely the CHMB \cite{ananthanarayan2020multiple}. This technique yields series solutions which, in the non-resonant case, are identical to those derived from the MoB. One must stress that at no step the convergence analysis of the basis series (or building blocks) is necessary to build the series solutions in the CHMB, whereas it is an unavoidable intermediate step for the MoB.

In the present work, we have taken up the task of revisiting the MoB. Among the important consequences of this study, we have illustrated that Rule 4 of MoB can lead to incomplete results. To explain this, a correspondence with the CHMB was proposed in Section \ref{Equivalence_Section}, in the nonresonant case, where only convergent series appear in the MoB. Extending this interconnection allows us to explain that divergent basis series that appear in the MoB only when higher-order poles are present in the corresponding MB representation and cannot be treated by using the heuristic expression Eq.\eqref{OldRule}. In the Appendix, we have also shown that the necessity of the \textit{ad-hoc} Rule 5 is due to the inadequacy of Rule 4 as, in the CHMB, the former rule is an obvious consequence of the residue theorem.

We end our discussion by briefly enumerating the reasons why the CHMB is more practicable than the MoB to evaluate Feynman integrals or certain classes of definite integrals.
\begin{enumerate}
     
    \item All integrals that are solvable by the MoB can be written as an MB representation. Therefore, such integrals can also be solved by applying the CHMB on the associated MB. 
    \item For cases where the basis series derived from MoB (or building blocks in the notation of CHMB) are higher-fold hypergeometric series, the convergence regions are challenging or, in some cases, impossible to derive. The CHMB circumvents this difficulty by finding the series representations, built as combinations of the building blocks, from a simple geometrical analysis.
    \item The emergence of termwise divergent basis series leads to incomplete results in the MoB, which never happens in the CHMB.
    \item The CHMB provides in general a master series for each solution, in the degenerate case, which is very helpful for numerical verification of the results.
    \item It is worth emphasizing that the CHMB has been automatized in \textit{Mathematica} in the \texttt{MBConicHulls.wl} package whereas, to our knowledge, there is no equivalent tool in the MoB case.
\end{enumerate}

Let us now briefly discuss the limitations shared by both methods. The convergence regions of the complete set of series solutions derived do not fill, in general, the entire domain of variables. The inaccessible regions, which can also include the boundaries of the convergence regions of the series solutions, are called ``white zones". An example of a white zone can be seen in the region of convergence plot of Fig.\ref{PlotROC}. As a corollary, when dealing with cases whose MB representation has a number of folds greater than the number of MB scales (the latter being $x_1,\cdots,x_N$ in Eq.\eqref{N_MB}) both methods generally fail, as the absence of a scale is equivalent to setting one of the $x_i$ to 1 in Eq.\eqref{N_MB}. This mostly falls in the white zones and the methods fail to find convergent series. The resolution of the white zones problem is an important and is left for future work.

Another interesting problem is to reformulate the MoB to get rid of the appearance of divergent series. An approach in this direction was discussed in \cite{gonzalez2017extension}, where an auxiliary variable was introduced in the intermediate steps to regularize the divergent series whose limit was taken to be zero in the final step. However, this is not a very useful strategy for complicated cases where more than one auxiliary variable may be necessary, implying that the final limiting value can be very hard to compute.

We finally note here that we have worked out several other Feynman integrals other than that presented in Section \ref{Box_Diagonal_Solve}, in order to confirm the statements made in this paper. Several definite integrals were also considered including the cases of the exponential integral function, the incomplete gamma function and the modified Bessel functions of second kind. 

\section{Acknowledgement}
We thank Sudeepan Datta for discussions at an early stage of this work.

\section*{Appendix A: {Obtaining Rule 5 from the MB analysis}} \label{Section_Rule_5}
In this appendix, we show that the heuristic Rule 5 (see section \ref{MoB_Rules}), introduced in \cite{gonzalez2016pochhammer} to solve possible incorrect evaluations of the MoB in particular situations, naturally emerges in the MB analysis which thus does not need to disentangle these situations from more common ones. Indeed, in \cite{gonzalez2016pochhammer}, it was shown from the study of entry 6.671.7 of \cite{gradshteyn2015table} that the evaluation of the definite integral in the LHS of Eq.(\ref{Deduce}) by MoB gives an incorrect $1/2$ factor in the RHS, for the $0<a<b$ case, and that a careful treatment of the Pochhammer symbols appearing in the intermediate steps of the calculations is necessary to get the correct result (this gave birth to Rule $5$ of MoB). We show here that this problem is intrinsic to MoB and that nothing wrong appears when using CHMB for the evaluation of this integral. To justify our claim, we evaluate the integral in Eq.(\ref{Deduce}) for $0<a<b$, from its MB representation,  and we show that the limiting value of Pochhammer symbols with negative index and negative integer base is properly taken into account by the conventional mathematics  of Cauchy's residue theorem.

The integral of entry 6.671.7 in \cite{gradshteyn2015table} reads
\begin{equation}\label{Deduce}
    I=\int_{0}^{\infty} J_{0}(ax) \mathrm{sin}(bx) \, dx=\left\{\begin{array}{ll}
{0} & {\text { if } 0<b<a} \\
{1 / \sqrt{b^{2}-a^{2}}} & {\text { if } 0<a<b}
\end{array}\right.
\end{equation}
where $J_0$ is a Bessel function of the first kind \cite{olver2010nist}.

In the MoB, one substitutes the power series expansion of $J_0$ and $\mathrm{sin}(x)$
\begin{equation}
J_{0}(a x)=\sum_{n_1=0}^{\infty} \phi_{n_1} \frac{a^{2 n_1}}{\Gamma(n_1+1) 2^{2 n_1}} x^{2 n_1}
\end{equation}
and
\begin{equation}
\sin (b x)=\sum_{n_2=0}^{\infty} \phi_{n_2} \frac{\Gamma(n_2+1)}{\Gamma(2n_2+2)} b^{2n_2+1} x^{2 n_2+1}
\end{equation}
 where $\phi_n=\frac{(-1)^n}{\Gamma(1+n)}$, in the LHS of Eq.\eqref{Deduce} and, by subsequently introducing the bracket symbol by Rule 3 of MoB, one obtains the following bracket series 
\begin{equation}\label{bracket66717}
I=\sum_{n_1,n_2=0}^{\infty}\phi_{n_1,n_2}\frac{\Gamma(n_2+1)\,a^{2n_1}\,b^{2n_2+1}}{2^{2n_1}\,\Gamma(2n_2+2)\Gamma(n_1+1)}\left\langle 2n_{12} +2 \right\rangle
\end{equation}
Similarly, using the MMoB, one can obtain the following MB representation
\begin{equation}\label{MB66717}
I=\frac{1}{2}\int\frac{d z_1}{2 \pi i}\,\,\frac{\Gamma(z^*_2+1)\,a^{2z_1}\,b^{2z^*_2+1}}{2^{2z_1}\,\Gamma(2z^*_2+2)\Gamma(z_1+1)}\Gamma(-z_1)\Gamma(-z^*_2)
\end{equation}
where, $z^*_2=-1-z_1$.
In \cite{gonzalez2016pochhammer}, precisely $n_1$ was taken as the free variable in Eq.(\ref{bracket66717}) to establish Rule 5. Hence, based on the connection discussed in Section \ref{Equivalence_Section} we consider in the MB representation of Eq.(\ref{MB66717}) the building block corresponding to the 1-combination of the gamma function $\Gamma(-z_1)$ whose poles are at $z_1=m_1$. This series can be obtained by closing the contour to the right. Therefore, using Cauchy's residue theorem, it is straightforward to show that the building block is
\begin{equation}\label{Deduce_1}
\begin{multlined}
    \frac{1}{2}\sum_{m_1=0}^{\infty} \textbf{Res}\Bigg[ \frac{\Gamma^2(-z_1-m_1)\,a^{2z_1+2m_1}\,b^{-1-2z_1-2m_1}}{2^{2z_1+2m_1}\Gamma(-2z_1-2m_1)}\Bigg]_{z_1=0}
    =\frac{1}{b\sqrt{\pi}}\sum_{m_1=0}^{\infty}\frac{\Gamma(m_1+1/2)}{m_1!}\left(\frac{a^2}{b^2}\right)^{m_1}\\
    =\frac{1}{b}\times{}_2F_1\left(\frac{1}{2},-,-;\frac{a^2}{b^2}\right)=\frac{1}{\sqrt{b^2-a^2}}
    \end{multlined}
\end{equation}
if $\vert\frac{a^2}{b^2}\vert<1$ which corresponds to the result shown in Eq.(\ref{Deduce}) for $0<a<b$.

One thus sees that there is nothing special in the calculation of this integral which, in fact, is straightforward to compute. Let us now show how the result of Rule 5, concerning Pochhammer symbols with negative argument and negative integer base, is implicit in this residue calculation.

For this we introduce the Pochhammer symbol as
\begin{equation}
    \frac{\Gamma(-z_1-m_1)}{\Gamma(-2z_1-2m_1)}=\frac{1}{(-z_1-m_1)_{-z_1-m_1}}
\end{equation}
and substitute it in the LHS of Eq.\eqref{Deduce_1} to get
\begin{equation}
\begin{multlined}
    \frac{1}{2}\sum_{m_1=0}^{\infty}\textbf{Res}\Bigg[ \frac{\Gamma(-z_1-m_1)\,a^{2z_1+2m_1}\,b^{-1-2z_1-2m_1}}{2^{2z_1+2m_1}(-z_1-m_1)_{-z_1-m_1}}\Bigg]_{z_1=0}
    \end{multlined}
\end{equation}
Now, applying the generalized reflection formula on $\Gamma(-z_1-m_1)$ and explicitly writing the residue using Cauchy's residue theorem, the following equivalent form is obtained
\begin{align}\label{Deduce_2}
     &\frac{1}{2}\sum_{m_1=0}^{\infty}(-1)^{m_1}\lim_{z_1\to 0}\Bigg[ \frac{a^{2z_1+2m_1}\,b^{-1-2z_1-2m_1}\Gamma(1-z_1)\Gamma(1+z_1)}{2^{2z_1+2m_1}{\Gamma}(1+z_1+m_1)(-z_1-m_1)_{-z_1-m_1}}\Bigg]\nonumber \\
     &=\frac{1}{2}\sum_{m_1=0}^{\infty}(-1)^{m_1}\frac{a^{2m_1}\,b^{-1-2m_1}}{2^{2m_1}\,\Gamma(1+m_1)}\lim_{z_1\to 0}\Bigg[ \frac{1}{(-z_1-m_1)_{-z_1-m_1}}\Bigg]
\end{align}
At this step, we see that the rule to take the limiting value of Pochhammer symbol naturally appears from the residue theorem. Indeed, the matching between Eq.(\ref{Deduce_2}) and the RHS of the first line of Eq. (\ref{Deduce_1}) implies
\begin{equation}\label{Pochhammer_Limit}
    \lim_{\epsilon \to 0} (-z_1-m_1))_{-z_1-m_1}=\frac{(-1)^{m_1} m_1 !}{\Gamma(2m_1+1)} \frac{1}{2}.
\end{equation}
which is nothing but Rule 5 (see Eq.(\ref{Rule_5}) for $k=1$). 

\section*{Appendix B: Proof of Identities}\label{Appendix_Identity}

In this section, we shall present the proofs of Eqs.\eqref{Identity1}, \eqref{Identity2} and \eqref{Identity3}, which are rewritten below:
\begin{align} 
1.& \, (B_{\sigma})^{-1} B_{\bar{\sigma}}+A^{\sigma}(A^{\bar{\sigma}})^{-1}=0 \label{IdentityA1}\\
2. & \, (B_{\sigma})^{-1}\,C-C'^{\, \sigma} +A^{\sigma}(A^{\bar{\sigma}})^{-1}\,C'^{\,\bar{\sigma}}=0 \label{IdentityA2} \\
3. &\,\, |\text{det}\, B_{{\sigma}_{1s}} |\,|\text{det} \, A^{\bar{\sigma}} |=|\text{det}\, B_{{\sigma}} | \label{IdentityA3}
\end{align}
Before we present the proof, let us recapitulate the involved matrices and notation.
\begin{equation}
B=\left(
\begin{array}{ccccc}
 b_{11} & b_{12} & \cdots & b_{1r} \\
 b_{21} & b_{22} & \cdots & b_{2r} \\
  & & \cdots & \\
 b_{s1} & b_{s2} & \cdots & b_{sr} \\
\end{array}
\right), \hspace{0.5cm} C=\left(
\begin{array}{ccccc}
c_1 \\
c_2 \\
. \\
 c_r \\
\end{array}
\right)\hspace{0.5cm}
A=\left(
\begin{array}{ccccc}
 B^{-1}_{{\sigma}_{1s}}\,B_{\bar{\sigma}_{1s}} \\\\
 -I_{r-s} \\
\end{array}
\right), \hspace{0.5cm}
C'=\left(
\begin{array}{ccccc}
B^{-1}_{{\sigma}_{1s}}\,C\\
 0_{r-s}\\
\end{array}
\right)
\end{equation}
where the matrices are of type $s\times r$ for $B$, $r \times 1$ for $C$, $r \times r-s$ for $A$ and $r \times 1$ for $C'$.
$\sigma$ denotes an arbitrary set of $s$ elements such that $\sigma \subset \{1,\cdots, r \}$ and $\bar{\sigma}=\{1,\cdots,r\} \setminus \sigma$. For brevity, we use $\sigma_{1s}=\{1,\cdots,s\}$ and $\bar{\sigma}_{1s}=\{1,\cdots,r\} \setminus \sigma_{1s}$. For any general $M$ matrix, we use the shorthand notation
\begin{equation}
M_{\sigma}= (M_{{\sigma}_1} \,  M_{{\sigma}_2} \,\, \cdots \,\, M_{{\sigma}_s}), \hspace{1cm} 
M^{{\sigma}}= \left(
\begin{array}{ccccc}
M^{{\sigma}_1} \\
M^{{\sigma}_2} \\
. \\
 M^{{\sigma}_s} \\
\end{array}
\right)
\end{equation}
where $M^{i}$ and $M_{j}$ denotes the $i^{th}$ row and $j^{th}$ column of matrix $M$. The mixed notation $M^{\{j_1 \,\, \cdots j_k\}}_{\{i_1 \,\, \cdots i_l\}} $ denotes the matrix constructed from the $\{ {j_1, \,\, \cdots, j_k} \}$ rows of the matrix $M_{\{i_1, \,\, \cdots, i_l\}}$.
\\
Let us now present our proofs.
\begin{enumerate}
\item \textbf{Proof of Eq.\eqref{IdentityA1}}: $(B_{\sigma})^{-1} B_{\bar{\sigma}}+A^{\sigma}(A^{\bar{\sigma}})^{-1}=0$\\\\
We start with the LHS which can be rewritten as,
\begin{align}
&(B_{\sigma})^{-1} B_{\bar{\sigma}}+A^{\sigma}(A^{\bar{\sigma}})^{-1}= (B_{\sigma})^{-1}\,(B_{\bar{\sigma}}A^{\bar{\sigma}}+B_{\sigma}A^{\sigma})\,(A^{\bar{\sigma}})^{-1} \nonumber \\
&(B_{\sigma})^{-1}\,(B_{\bar{\sigma}}A^{\bar{\sigma}}+B_{\sigma}A^{\sigma})\,(A^{\bar{\sigma}})^{-1}=(B_{\sigma})^{-1}\,\left(\sum_{i} B_{\bar{\sigma_{i}}}A^{\bar{\sigma_{i}}}+\sum_{j}B_{\sigma_j}A^{\sigma_j}  \right)\,(A^{\bar{\sigma}})^{-1}
\end{align}
We note that $\sigma \cup \bar{\sigma}=\{1,\cdots,r\}$. Therefore, the term inside the parenthesis is same as $\sum_{i\in \{1, \cdots , r \}} B_i A^i= B A$. Now, from the definition of $B$ and $A$ we have
\begin{equation}
B A= \left( B_{\sigma_{1s}} \,\, B_{\bar{\sigma}_{1s}} \right) \left(
\begin{array}{ccccc}
 B^{-1}_{{\sigma}_{1s}}\,B_{\bar{\sigma}_{1s}} \\\\
 -I_{r-s} \\
\end{array}
\right)=0
\end{equation} 
Therefore,
\begin{equation}
(B_{\sigma})^{-1} B_{\bar{\sigma}}+A^{\sigma}(A^{\bar{\sigma}})^{-1}=(B_{\sigma})^{-1}\,BA\,(A^{\bar{\sigma}})^{-1}=0
\end{equation}
which proves the identity.
\item \textbf{Proof of Eq.\eqref{IdentityA2}}: $(B_{\sigma})^{-1}\,C-C'^{\, \sigma} +A^{\sigma}(A^{\bar{\sigma}})^{-1}\,C'^{\,\bar{\sigma}}=0$\\\\
We apply the above identity to rewrite the R.H.S as,
\begin{align}
(B_{\sigma})^{-1}\,C-C'^{\, \sigma} +A^{\sigma}(A^{\bar{\sigma}})^{-1}\,C'^{\,\bar{\sigma}}
&=(B_{\sigma})^{-1}\,C -C'^{\, \sigma} -(B_{\sigma})^{-1} B_{\bar{\sigma}}\,C'^{\,\bar{\sigma}}
\nonumber \\
&=(B_{\sigma})^{-1}\,C-C'^{\, \sigma}-(B_{\sigma})^{-1} \left( B_{\bar{\sigma}}\,C'^{\,\bar{\sigma}} + B_{\sigma} C'^{\,\sigma} - B_{\sigma} C'^{\,\sigma} \right) \nonumber \\
&=(B_{\sigma})^{-1}\,C-C'^{\, \sigma}-(B_{\sigma})^{-1} \left( B\,C' - B_{\sigma} C'^{\,\sigma} \right) \nonumber \\
&=(B_{\sigma})^{-1}\,C-(B_{\sigma})^{-1} B\,C' \nonumber \\
&=(B_{\sigma})^{-1}\,C-(B_{\sigma})^{-1} B_{\sigma_{1s}}\,C'^{\, \sigma_{1s}} \nonumber \\
&=(B_{\sigma})^{-1}\,C-(B_{\sigma})^{-1} B_{\sigma_{1s}}\,B^{-1}_{\sigma_{1s}}\,C \nonumber \\
&=(B_{\sigma})^{-1}\,C-(B_{\sigma})^{-1}\,C=0
\end{align}
which completes the proof.
\item \textbf{Proof of Eq.\eqref{IdentityA3}}: $|\text{det}\, B_{{\sigma}_{1s}} |\,|\text{det} \, A^{\bar{\sigma}} |=|\text{det}\, B_{{\sigma}} | $\\\\
To ease the proof, we first rewrite matrix $B$ as
\begin{equation}
B=\left( B_1 \,\, B_2 \,\,\cdots \,\,B_s\right)\,\left( I_s \,\, B'_{s+1} \,\, \cdots \,\, B'_r\right)= B_{\{1, \, \cdots\, , s\}}\, B'=B_{\sigma_{1s}}\,B'
\end{equation}
where $B'_i=B^{-1}_{\sigma_{1s}}\,B_i$ and the new matrix $B'=\left( I_s \,\, B'_{s+1} \,\, \cdots \,\, B'_r\right)$ is of type $s\times r$.
From this definition of $B'$, we have

\begin{equation}
B_{\{i_1, \cdots, i_l\}}=B_{\sigma_{1s}}\,B'_{\{i_1, \cdots, i_l\}}, \hspace{1cm}
A=\left(
\begin{array}{ccccc}
 B^{-1}_{{\sigma}_{1s}}\,B_{\bar{\sigma}_{1s}} \\\\
 -I_{r-s} \\
\end{array}
\right)= \left(
\begin{array}{ccccc}
 B'_{\bar{\sigma}_{1s}} \\\\
 -I_{r-s} \\
\end{array}
\right)
\end{equation}
Next, we decompose $\bar{\sigma}$ as $\bar{\sigma}=\sigma_1\cup \sigma_2$, such that,
\begin{equation}
\sigma_1=\bar{\sigma} \setminus \{s+1, \cdots \,, r\}, \hspace{1cm} \sigma_2=\bar{\sigma}\cup \{s+1, \cdots \,, r\}
\end{equation}
so $\sigma_1\subset \{1, \cdots, s\}$ and $\sigma_2\subset \{s+1, \cdots, r\}$. Hence, we can write $A^{\bar{\sigma}}$ as
\begin{equation}
A^{\bar{\sigma}}= \left(
\begin{array}{ccccc}
  A^{\sigma_1} \\\\
  A^{\sigma_2} \\
\end{array}
\right)= \left(
\begin{array}{ccccc}
 B'^{\,\sigma_1}_{\bar{\sigma}_{12}} \\\\
   -I^{\,\sigma_2}_{r-s}  \\
\end{array}
\right)
\end{equation}
where $B'^{\,\sigma_1}_{\bar{\sigma}_{12}}$ denotes the matrix constructed from the rows $\sigma_1$ of matrix $B'_{\bar{\sigma}_{1s}}$.
\\
Therefore, we can write $|\text{det}(A^{\bar{\sigma}})|$ as
\begin{equation}
|\text{det}(A^{\bar{\sigma}})|= \left|\text{det}\left(
\begin{array}{ccccc}
   B'^{\sigma_1}_{\bar{\sigma}_{12}}  \\\\
  -I^{\,\sigma_2}_{r-s}  \\
\end{array}\right)
\right|= \left|\text{det}\left(
\begin{array}{ccccc}
   B'^{\,\sigma_1}_{\sigma_3}  \\
\end{array}\right)
\right|
\end{equation}
where we simplified the determinant by evaluating it along the rows of $-I^{\,\sigma_2}_{r-s}$ and by defining a new set $\sigma_3$ as follows
\begin{align}
\sigma_3 &=\{r+1, \cdots, s\} \setminus \sigma_2=\{r+1, \cdots, s\}\setminus \bar{\sigma} \nonumber \\
&=\{r+1, \cdots, s\} \setminus \left( \{1, \cdots, r \} \setminus \sigma \right)  \nonumber \\
&=(\sigma\cap \{r+1, \cdots, s\}) \cup (\{r+1, \cdots, s\} \setminus \{1, \cdots, r \}) \nonumber \\
&=\sigma\cap \{r+1, \cdots, s\}
\end{align}
Here we see that $\text{det}\left(
\begin{array}{ccccc}
   B'^{\,\sigma_1}_{\sigma_3}  \\
\end{array}\right)$ will remain unchanged if we append the matrix $(I_{s})_{\sigma_4}$ to the right and $ B'^{\,\sigma_4}_{\sigma_3} $ below of $B'^{\,\sigma_1}_{\sigma_3}$, for $\sigma_4=\{1,\cdots,s\} \setminus \sigma_1=\sigma \cap \{1,\cdots,s\}$. Thus, we have
\begin{equation}
|\text{det}(A^{\bar{\sigma}})|=\left|\text{det}\left(
\begin{array}{ccccc}
   B'^{\,\sigma_1}_{\sigma_3} & &  \\
   & & (I_{s})_{\sigma_4} \\
   B'^{\,\sigma_4}_{\sigma_3} & & \\
\end{array}\right)
\right|
\end{equation}
We note that $\sigma_1 \cup \sigma_4 =\{1,\cdots,s\}$, thus $\left(
\begin{array}{ccccc}
   B'^{\,\sigma_1}_{\sigma_3}\\
   B'^{\,\sigma_4}_{\sigma_3}\\
\end{array}\right)=\left( B'_{\,\sigma_3}\right)$. Moreover, by definition $\sigma_4 \subset \{1,\cdots, s\}$, hence we can write $(I_{s})_{\sigma_4}=B'_{\sigma_4}$, so
\begin{equation}
|\text{det}(A^{\bar{\sigma}})|=\left|\text{det}\left(
B'_{\,\sigma_3}\,\, B'_{\,\sigma_4}\right)
\right|
\end{equation}
Finally, as $\sigma_3 \cup \sigma_4=\sigma$, we have $(B'_{\,\sigma_3}\,\, B'_{\,\sigma_4})=B'_{\,{\sigma}} $. Therefore,
\begin{align}
|\text{det}(B_{\sigma_{1s}})||\text{det}(A^{\bar{\sigma}})|&=|\text{det}(B_{\sigma_{1s}})||\text{det}(B'_{{\sigma}})| \nonumber \\
&=|\text{det}(B_{\sigma_{1s}} \, B'_{{\sigma}})| \nonumber \\  
&=|\text{det}(B_{\sigma})|
\end{align}
which completes the proof.
\end{enumerate}

\bibliography{equi} 
\bibliographystyle{unsrt}
\end{document}